# Synesthetic Dice: Sensors, Actuators, And Mappings

Albrecht Kurze

Chemnitz University of Technology, Albrecht.Kurze@informatik.tu-chemnitz.de

How bright can you cry? How loud does the sun shine? We developed a multisensory and multimodal tool, the *Loaded Dice*, for use in co-design workshops to research the design space of IoT usage scenarios. The Loaded Dice incorporate the principle of a technical synesthesia, being able to map any of the included sensors to any of the included actuators. With just a turn of one of the cubical devices it is possible to create a new combination. We discuss the core principles of the *Loaded Dice*, what sensors and actuators are included, how they relate to human senses, and how we realized a meaningful mapping between sensors and actuators. We further discuss where we see additional potential in the *Loaded Dice* to support synesthetic exploration – as *Synesthetic Dice* – so that you can eventually find out who cries brighter.

## 1 INTRODUCTION

Some years ago we designed and developed the *Loaded Dice* [8,9], a multisensory and multimodal hybrid toolkit to ideate Internet of Things (IoT) devices and scenarios, e.g. for the 'smart' home, and with different groups of co-designers [3,7,8]. The *Loaded Dice* filled a gap between analog, non-functional tools, often card-based, e.g. *KnowCards* [1], and functional but tinkering based tools, e.g. *littleBits* [2], for multisensory und multimodal exploration, ideation and prototyping.

We introduce the *Loaded Dice*, the core concepts that they are built on, the used sensors and actuators, and how they map to different human senses. We will then continue to discuss how we realized mappings between sensed raw value, normalized intermediate values, and actuated values. While the mappings that we currently use are sufficiently good enough for current purposes, we see big potential in some extended uses as *'Synesthetic Dice'*.

This brings us to our core question: *How can the Loaded Dice be used for exploration and research of synesthetic mappings between sensors and actuators, e.g. for innovative interactions and non-verbal communication?*

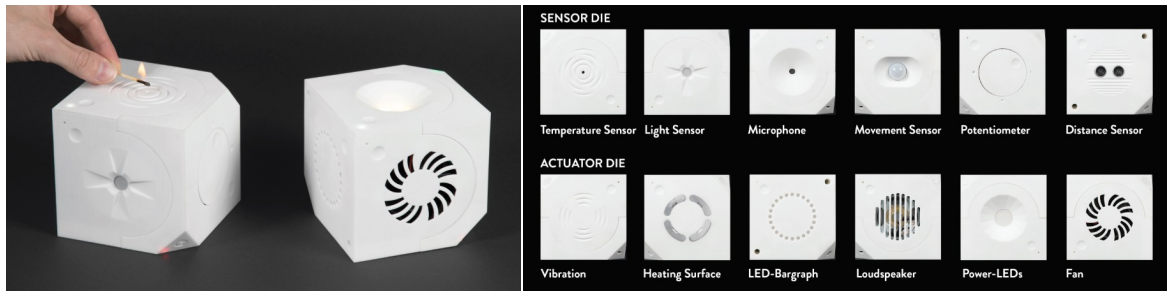

Figure 1: The Loaded Dice; left: example of devices in use, turning heat into light (sensor die with temperature sensor active and actuator die with power LED active) [9]; right faces and functions – sensors and actuators [8]

## 2 THE LOADED DICE - SENSES, SENSORS, ACTUATORS

The *Loaded Dice*[1] are a set of two cubical devices wirelessly connected (fig. 2a). Each cube has six sides, offering in one cube six sensors and in the other cube six actuators, one on each side, suitable for multisensory and multimodal environmental and user interactions. The sensor cube normalizes a raw sensor value meaningfully, transmits it, and then the other cube actuates it mapped on an output. The cubical shape communicates the intuitive reading that the top side is active, like a die, offering an easy and spontaneous way to re-combine sensors and actuators. Every sensor-in and actuator-out combination is possible resulting in 36 combinations in total. [5]

The "traditional five" human senses are sight, hearing, taste, smell and touch. Secondary senses are temperature, pain, proprioception and balance. Due to the constraints of the technical platform we could not address all human senses with sensors and actuators. Overall the Loaded Dice holds sensors and actuators equivalent to some human senses directly (see fig. 1 and table 1 for details). It is also possible to think about effects to address other senses using the given sensors and actuators, e.g. to inflict pain via the Peltier element through excessive heat or cold (not intended nor recommended). It is also possible (but currently not implemented) to use the internal inertial measurement unit (IMU), consisting of an accelerometer and gyrometer, not only for interaction controls but also as a sense, as an equivalent to proprioception and balance (movement and position).

New multisensory interaction modalities are possible but not yet implemented, e.g. olfactory / smell. They have the potential to broaden interaction qualities even further and especially in an emotional way [6].

Table 1. Human senses vs. sensors and actuators in the Loaded Dice

| Human Sense | Sensor | Actuator |
| --- | --- | --- |
| sight (visual stimuli) | luxmeter (visible light luminosity/ brightness) passive infrared detector (PIR movement) ultrasonic transceiver (distance) | power LED (brightness) LED ring-graph (count, overall brightness, color) |
| hearing (auditive stimuli) | microphone (amplitude) | sound (modulated note for instrument) (vibration motor, rattling noise) (fan, air flow noise) |
| touch (tactile stimuli) | potentiometer (manual angular dial of 270°) | vibration motor (vibration) fan (mechanical stimulation on hairs) |
| temperature (thermal stimuli) | infrared thermometer (thermopile / thermal radiation) | Peltier element (cooling and heating plate) fan (cooling by chill effect on skin) |

---

[1] video demonstrating the Loaded Dice: https://www.youtube.com/watch?v=-E5aUiktCic

## 3 SYNESTHESIA - MAPPING SENSES AND MODALITIES

Synesthesia describes the phenomenon of an event being experienced by another, separate sensory modality [4]. While medical not exact, in principle, this means a sound might not only be heard but also be seen as a color (as an example). Most existing tools, i.e. for IoT ideation, do not employ synesthesia effects as a design opportunity in order to break with existing sensing stereotypes for framing design spaces. Such a stereotype could be e.g. that making noise should always be connected with hearing noise. While most related digital (IoT) ideation tools do allow for flexible combinations of different sensors and actuators in principle, this is not ad hoc possible. Instead they require necessary steps in combining parts or mapping sensor values to actuator values. Thus, they demand an initial idea of how the combination should play out. Our tool allows users to explore such synesthetic effects ad hoc.

We implemented a meaningful mapping between every sensor and actuator that is used in the Loaded Dice. This includes reasonably chosen sampling rates, ranges and steppings for raw input values, their normalization on internal values and the conversion back to meaningful output values. All this is done internally in hard- and software, without the need of user intervention. Selecting a new sensor-actuator combination just requires bringing another side to the top. Based on the presented design rationale, a co-designer can transport heat over a distance by choosing the infrared thermometer and Peltier element sides of both cubes. Rotating the actuator cube to the power-LEDs would transform the temperature into light, thus mimicking synesthesia-like perception.

The possibilities of the Loaded Dice can be used in a framed scenario-driven co-design approach, in open exploration or even just for 'sensory sketching', even for 'weird' synesthetic combinations, e.g:

- to try out how bright sunlight sounds or feels as vibration
- what temperature a loud cry has
- how much air-flow half a meter distance is
- whether you can feel the flickering of light …

We use meaningful but simple functions for preprocessing of raw sensor values and normalization to an intermediate data value as well back to actuations (table 2). Overall, the mappings are done in a predefined 'static' way. However, static does not mean one fits all. It is necessary to consider non-linearities and dynamics, e.g. for light and sound, as these senses are not perceived in a linear or static manner by humans. However, we applied 'just good enough' assumptions for meaningfulness without the claim of physical or psychometric correctness, sometimes even a bit off to make effects clearer. Currently also the sensor as well as the selected actuator are considered for the mapping in addition to the normalized value. We do this mainly for technical reasons as the different modalities operate at different speeds. Currently, only the LED ring graphic signals which sensor has sampled the data by changing color.

Table 2: Current mapping from sensed values to intermediate values and then to actuated values

| Sensor | Sensor Mapping | Value | Actuator Mapping | Actuator |
|---|---|---|---|---|
| potentiometer | 0..270° AD sampling 0..1023 → linear → 0..24 | | → Neopixels count 0..24, color coded by sensor, brightness per pixel static | ring-graph |
| thermometer | digital read-out 0..50 ℃ → linear → 0..24 | | → sqr → 0..576 → 0..255 RGB brightness | power LED |
| microphone | 50ms window AD sampling 0..1023 → max-min difference → 0..1023 → linear → 0..24 | 0..24 | 0 → 0; 1..24 → MIDI noteOn(value+50) | sound |
| distance | 0 → 0; 1..72 cm → linear → 1..24 | | 0..12 → ·-255..0 (cooling) 12..24 → ·0..255 (heating) PWM or 0..24 → 0..255 (from neutral to heating only) PWM | Peltier thermo |
| PIR movement | binary 0 → 0; 1 → 24 | | 0 → 0; 1..24 → 64..255 PWM | vibration |
| light | digital read-out 0..65535 lx →·sqrt →·0..48 → 0..24 | | 0 → 0; 1..24 → 160..255 PWM | fan |

Every combination is possible, alignment in lines just as examples. AD: analog→digital conversion, PWM: pulse width modulation

While we are quite satisfied what the Loaded Dice can already do there are some new possibilities at hand:
- more use of colors: for power LED element and LED ring-graph (NeoPixels are colorful…)
- other use of sound: other (music/midi) instruments, modulation of velocity and pitch, other sounds (artificial or sampled in nature)
- use of spatial component: position of the LEDs of the ring-graph, color fades, patterns
- use of temporal components: from time static value to dynamic patterns for sound, vibration, light, air flow etc.

A flexible "sketching" of a new mapping function would allow to bring in completely new synesthesia effects, also not necessarily only limited to one input sensor and one output actuator at one time.

## 4 CONCLUSION

While the Loaded Dice can already be used meaningfully for activities associated with synesthesia, e.g. for ideation, we see a lot of potential in more flexible mappings and even other creative uses of what the sensors and actuators might do. We are open for inspirations and ideas.

## ACKNOWLEDGMENTS

This research is funded by the German Ministry of Education and Research (BMBF), grant FKZ 16SV7116.